\documentclass[twocolumn,prl]{revtex4}
\usepackage{amsmath,amssymb,bm}
\usepackage{graphicx}
\usepackage{epstopdf}
\usepackage{latexsym}
\usepackage{subfigure}
\usepackage[usenames,dvipsnames]{color}
\usepackage{hyperref}
\usepackage{natbib}

\newcommand {\be}{\begin{equation}}
\newcommand {\ee}{\end{equation}}
\newcommand {\ba}{\begin{eqnarray}}
\newcommand {\ea}{\end{eqnarray}}

\begin{document}

\title{Microscopic model for Feshbach interacting fermions in an optical lattice with arbitrary scattering length and resonance width}
\author{M. L. Wall$^1$ and L. D. Carr$^{1,2}$}
\address{$^1$Department of Physics, Colorado School of Mines, Golden, CO 80401, USA}
\address{$^2$Universit\"at Heidelberg, Physikalisches Institut, D-69120 Heidelberg, Germany}

\begin{abstract}
We numerically study the problem of two fermions in a three dimensional optical lattice interacting via a zero-range Feshbach resonance, and display the dispersions of the bound states as a two-particle band structure with unique features compared to typical single-particle band structures.  We show that the exact two-particle solutions of a projected Hamiltonian may be used to define an effective two-channel, few-band model for the low energy, low density physics of many fermions at arbitrary $s$-wave scattering length.  Our method applies to resonances of any width, and can be adapted to multichannel situations or higher-$\ell$ pairing.  In strong contrast to usual Hubbard physics, we find that pair hopping is significantly altered by strong interactions and the presence of the lattice, and the lattice induces multiple molecular bound states.
\end{abstract}

\maketitle

The crossover of a system of attractive two-component fermions from a condensate of loosely bound Cooper pairs to a condensate of tightly bound bosonic molecules has a long history~\cite{1-2}, and appears in many contexts, including high-temperature superconductivity~\cite{Ranninger_Robin_95} and ultracold atoms~\cite{Kohl_Moritz_05}.  Furthermore, near the crossover such a system enters the \emph{unitary regime} where the scattering length is larger than any other length scale in the problem.  The physics of this regime is relevant to many different fields, bringing together quantum chromodynamics, holographic duality, and ultracold quantum gases~\cite{focus}.  Theoretical study of the unitary regime is generally difficult due to the absence of any small parameter.

Theoretical analysis becomes even more difficult in a lattice, as the center of mass, relative, and internal degrees of freedom become coupled, leading to composite particles whose properties depend on their center of mass motion~\cite{Cafolla_Schnatterly_85}.  Furthermore, strong interactions require the inclusion of a large number of Bloch bands for an accurate description, and this cannot be handled efficiently by modern analytical or numerical many-body techniques.  In addition to general theoretical interest in how fermions pair to form bosons in a discrete lattice setting, the study of pairing in lattices is of significant practical importance.  For example, an accurate, systematically correctable, and computationally feasible many-body Hamiltonian is necessary for calibrating fermionic quantum simulators as has been done in the bosonic case~\cite{Trotzky_Pollet_10}.

\begin{figure}[b]
\vspace{-0.1in}
\centerline{\includegraphics[width=\columnwidth]{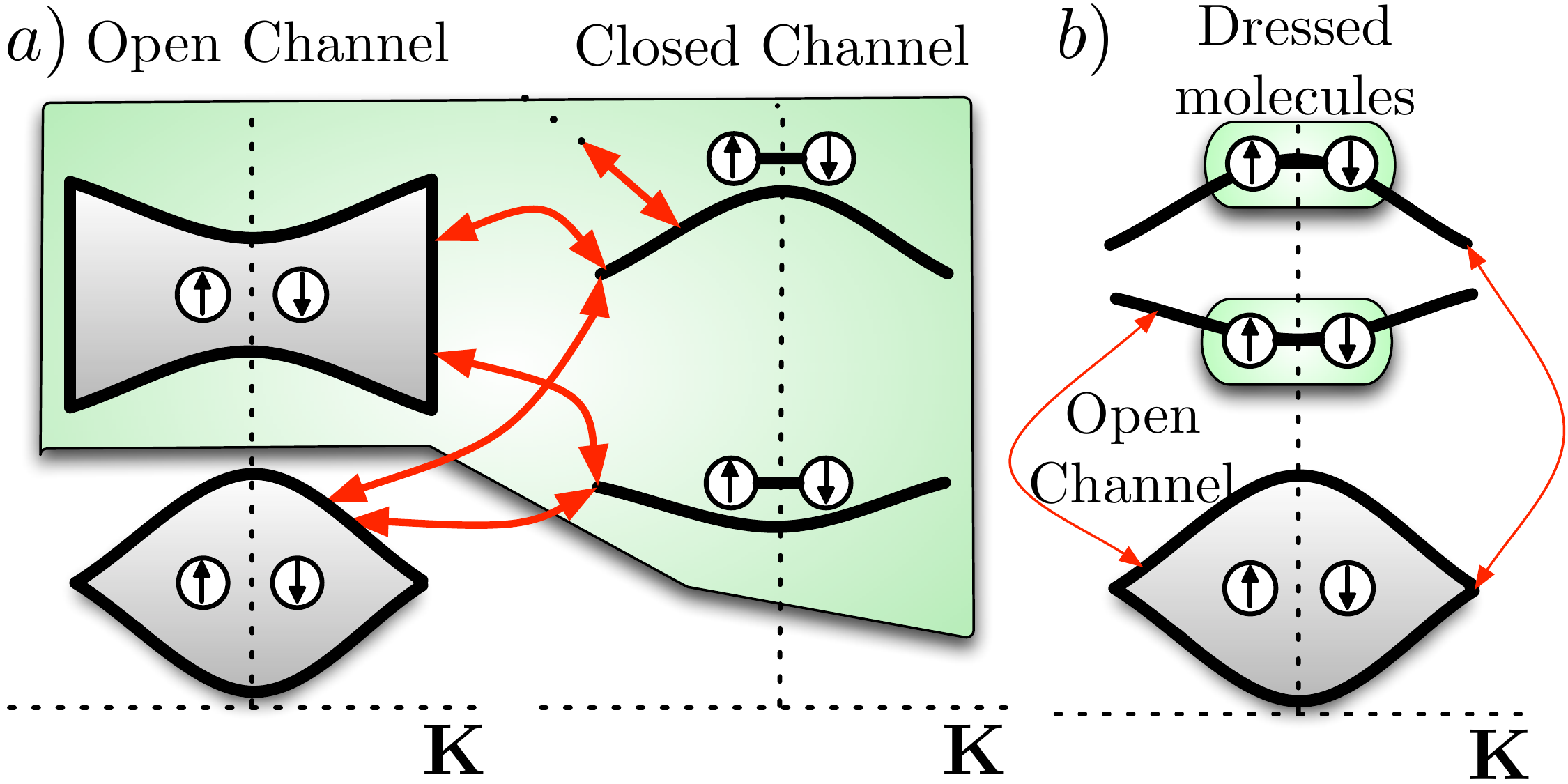}}
\caption{\label{fig:EMT} (Color online)  \emph{Schematic of the FRH transformation}. (a) In a broad Feshbach resonance, all two-particle scattering continua (gray shading) are strongly coupled to bare molecular bands (solid lines).  Thus all scattering continua are virtually strongly coupled.  (b) By correctly dressing the molecular bands, one obtains a single scattering continuum (gray) plus well-separated dressed molecular bands (green), with much simpler couplings.  This is our efficient, numerically tractable, FRH.
}
\end{figure}

In this Letter, we describe a general method to derive an effective few-band low-energy Hamiltonian for Feshbach interacting fermions in a lattice from the numerical solution of the two-body problem.  We call this Hamiltonian the \emph{Fermi Resonance Hamiltonian} (FRH).  This method applies to Feshbach resonances of any width and for arbitrary scattering length, and all parameters appearing in the effective model can be computed microscopically from the properties of the two-body solution.  The difference between the bare model and the FRH is sketched in Fig.~\ref{fig:EMT}.

%

The simplest approach to describing Feshbach interacting fermions is to replace the interaction with a pseudo-potential chosen to reproduce the correct scattering length.  When restricted to a single Bloch band, this leads to the popular Hubbard model~\cite{Esslinger_10} which has been shown to break down for scattering lengths which are far from being resonant, even when the parameters appearing in the model are determined self-consistently from few-body physics~\cite{Buechler_10+}.  Our work instead defines a ``dressed" closed channel whose properties are chosen to reproduce both the scattering and bound states correctly.  In contrast to past two-channel approaches~\cite{10-14}, we construct the dressed fields using the full lattice solution and not an approximation where the center of mass and relative coordinates separate, such as the harmonic oscillator potential.  The use of any separable approximation leads to qualitative errors, such as the lack of tunneling along non-principal axes, and quantitative errors, such as the underestimation of principal axis tunneling matrix elements, often by an order of magnitude.

\emph{Exact solution for two particles.} The basic concept of a two-channel model is for an \emph{open} channel to describe scattering between two atoms and a separate \emph{closed} channel to describe bound pairs.  While each channel represents a single scattering or bound state in the continuum, in the lattice it also acquires a band index.  Because of an inter-channel coupling, the actual molecule is a superposition of bands from both channels.

To treat this problem, we begin with the nonlinear eigenvalue problem developed by B\"uchler~\cite{Buechler_10+} for $E_{\mathbf{K}}$, the bound state energy at total quasimomentum $\mathbf{K}$, and $R_{\mathbf{s}}^{\mathbf{K}}$, the coefficients of the closed channel portion of the wavefunction.  As shown in~\cite{Buechler_10+,SupplMat}, for the bound states of two fermions in an optical lattice interacting via a zero-range Feshbach resonance in a two-channel model:
\begin{align}
\label{eq:NLEE}&\left[E_{\mathbf{K}}-\nu-{E}_{\mathbf{s}\mathbf{K}}^M\right]R_{\mathbf{s}}^{\mathbf{K}}=\textstyle{\frac{g^2}{a^3}}\textstyle{\sum_{\mathbf{t}}}\chi_{\mathbf{st}}^{\mathbf{K}}\left(E_{\mathbf{K}}\right)R_{\mathbf{t}}^{\mathbf{K}}\, ,\\
\label{eq:NLEE2}&\chi_{\mathbf{st}}^{\mathbf{K}}\left(E_{\mathbf{K}}\right)\equiv\textstyle{\int\frac{d\mathbf{q}}{v_0}\sum_{\mathbf{m}\mathbf{n}}}\frac{h_{\mathbf{s}\mathbf{K}}^{\mathbf{nm}}\left(\mathbf{q}\right) {h_{\mathbf{t}\mathbf{K}}^{\mathbf{nm}}}^{\star}\left(\mathbf{q}\right)}{E_{\mathbf{K}}-E_{\mathbf{nm}}^{\mathbf{K}}\left(\mathbf{q}\right)+i\eta}-\bar{\chi}_{\mathbf{st}}^{\mathbf{K}}\, ,\\
\label{eq:NLEE3}&\bar{\chi}_{\mathbf{st}}^{\mathbf{K}}\equiv-\textstyle{\int\frac{d\mathbf{q}}{v_0}\sum_{\mathbf{m}\mathbf{n}}}\bar{h}_{\mathbf{s}\mathbf{K}}^{\mathbf{nm}}\left(\mathbf{q}\right) {\bar{h}_{\mathbf{t}\mathbf{K}}^{\mathbf{nm\star}}}\left(\mathbf{q}\right)/\bar{E}_{\mathbf {nm}}^{\mathbf{K}}\left(\mathbf{q}\right)\, .
\end{align}
Here $\nu$ is the renormalized detuning between the open and closed channels, $g$ is the inter-channel coupling, $a$ is the lattice spacing, $v_0$ is the volume of the Brillouin zone (BZ), and the bars in Eqs.~(\ref{eq:NLEE2}-\ref{eq:NLEE3}) denote quantities computed in the absence of an optical lattice.  We assume that spin-spin interactions which change the orbital angular momentum are negligible so that the scattering is purely $s$-wave.  The optical lattice is assumed to be simple cubic with lattice spacing $a$ and potential $V\left(\mathbf{x}\right)=V\sum_{j\in\left\{x,y,z\right\}}\sin^2\left(\pi j /a\right)$.  The overlaps of the dimensionless coupling between the open and closed channels are
\begin{align}
\nonumber\frac{h_{\mathbf{s}\mathbf{K}}^{\mathbf{nm}}\left(\mathbf{q}\right)}{\sqrt{N^3a^3}}&=\textstyle{\int }d\mathbf{x}d\mathbf{y}\left[\psi_{\mathbf{nq}}\left(\mathbf{x}\right)\psi_{\mathbf{m},\mathbf{K}-\mathbf{q}}\left(\mathbf{y}\right)\right]^{\star}\alpha\left(\mathbf{r}\right)\phi_{\mathbf{sK}}\left(\mathbf{R}\right)\, ,
\end{align}
where $N^3$ is the number of unit cells, the $\psi_{\mathbf{n}\mathbf{q}}\left(\mathbf{x}\right)$ are Bloch functions with energies $E_{\mathbf{nq}}$ for particles with mass $m$ spanning the open channel and $\phi_{\mathbf{sK}}\left(\mathbf{z}\right)$ are Bloch functions with energies $E^M_{\mathbf{sK}}$ for particles with mass $2m$ in a lattice potential $2V$ spanning the closed channel.  We have also defined relative $\mathbf{r}\equiv\mathbf{x}-\mathbf{y}$ and center of mass $2\mathbf{R}=\mathbf{x}+\mathbf{y}$ coordinates, and $\alpha\left(\mathbf{r}\right)$ is a regularization of the inter-channel coupling.  The sum of the noninteracting energies of the open channel is denoted $E_{\mathbf{nm}}^{\mathbf{K}}\left(\mathbf{q}\right)=E_{\mathbf{nq}}+E_{\mathbf{m},\mathbf{K}-\mathbf{q}}$ and the zero of energy is $E_{\mathbf{11}}^{\mathbf{0}}\left(\mathbf{0}\right)$.  Here and throughout the rest of this work $\mathbf{n}$ and $\mathbf{m}$ are band indices for the open channel, $\mathbf{s}$ and $\mathbf{t}$ are band indices for the closed channel, $\mathbf{q}$ is a single-particle quasimomentum, and $\mathbf{K}$ is the total quasimomentum.

%
\begin{figure}[t]
\centerline{\includegraphics[width=0.8\columnwidth]{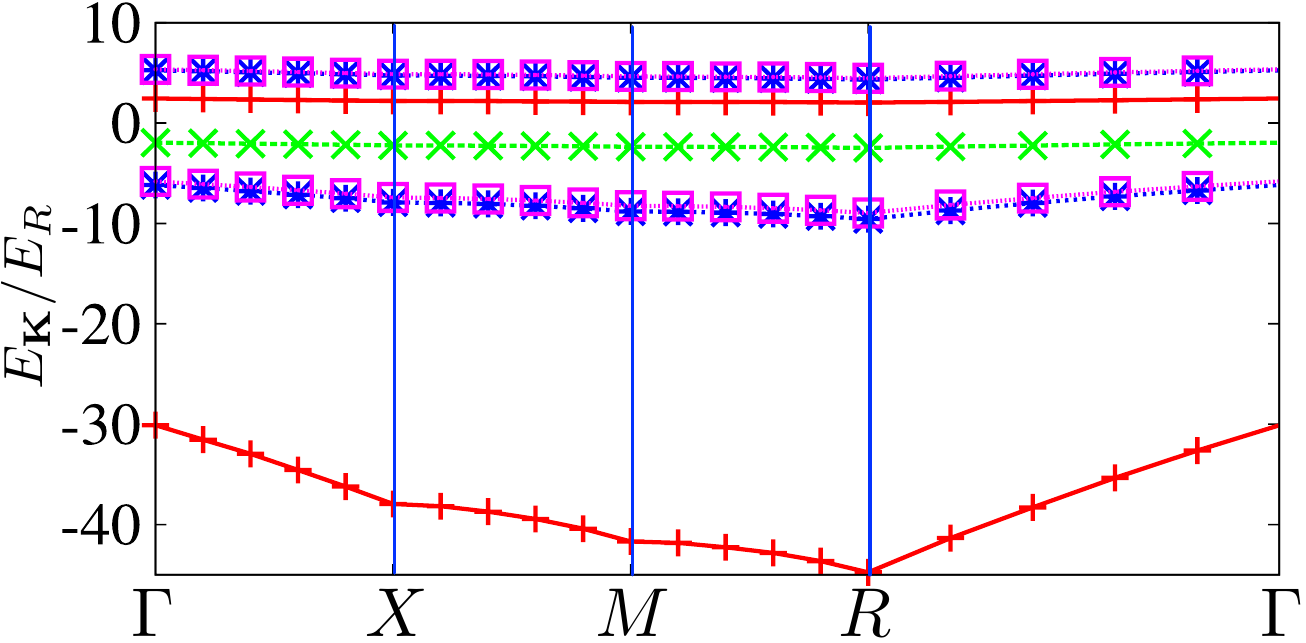}}
\caption{\label{fig:IsoBS} (Color online)  \emph{Exact two-particle band structures for various $a_s$ in a strong optical lattice}.  The bound state energies for $a_s/a=-5$ (purple boxes), $-0.1$ (red pluses), $0.1$ (green crosses), and $5$ (blue asterisks) as a function of the total quasimomentum $\mathbf{K}$ along a path connecting the high-symmetry points in the irreducible BZ $\Gamma=\left(0,0,0\right)$, $X=\left(-\pi/a,0,0\right)$, $M=\left(-\pi/a,-\pi/a,0\right)$, $R=\left(-\pi/a,-\pi/a,-\pi/a\right)$ for a lattice with $V/E_R=12$.  The near-resonant points $a_s/a=\pm 5$ lie nearly on top of one another, demonstrating universality.}
\vspace{-0.2in}
\end{figure}

While Eqs.~(\ref{eq:NLEE})-(\ref{eq:NLEE3}) apply to resonances of any width, we focus on the experimentally relevant limit of a broad resonance.  Narrow resonances are treated in the supplementary material~\cite{SupplMat}. A broad resonance in the few-body sense is the limit of effective range much smaller than the lattice spacing, $r_B\ll a$, and so we can take the limits $g/E_Ra^{3/2}=4\sqrt{a/r_B\pi^3}\to \infty$, $\nu/E_R\to\infty$, $a_s/a=-\pi g^2/8 a^3E_R \nu = \mathrm{const.}$, to obtain $(8a_sE_R/\pi a)\chi^{\mathbf{K}}\left(E_{\mathbf{K}}\right)\mathbf{R}^{\mathbf{K}}-\mathbf{R}^{\mathbf{K}}=0,$ where $E_R=\hbar^2\pi^2/2ma^2$ is the recoil energy and $a_s$ is the $s$-wave scattering length.  How can we then obtain the dispersion relation $E_{\mathbf{K}}$ for fixed $a_s$, $\nu$, etc.?  First, fix the energy eigenvalue $E_{\mathbf{K}}$ and solve the resulting \emph{linear} eigenproblem for $1/a_s$.  This provides exact eigentuples $\left(E_{\mathbf{K}},a_s,\mathbf{R}^{\mathbf{K}}\right)$ of the nonlinear eigenproblem, though it may not be the $a_s$ we seek.  Second, fix $a_s$ and use the exact tuple nearest this value as initialization for a Newton-Raphson iteration~\cite{lancaster_66}.  This two-stage approach converges to a relative accuracy of 0.01\% in a few tens of iterations~\cite{wallforthcoming}.

Because the eigenequation Eq.~\eqref{eq:NLEE} is invariant under translation by any Bravais lattice vector, its eigenvalues can be classified according to the total quasimomentum and shown as a two-particle band structure.  A complete classification of the solutions is given in the supplemental material~\cite{SupplMat}.  In Fig.~\ref{fig:IsoBS} we show only the energy of the low-energy bound states with completely even parity under inversion as a function of $\mathbf{K}$ for several $a_s/a$ in a lattice with $V/E_R=12$.  We see the appearance of several bound states for a fixed $s$-wave scattering length, in contrast to the continuum.  These additional bound states arise from the coupling of quasimomenta modulo a reciprocal lattice vector induced by the reduced translational symmetry.  One salient feature is the emergence of universality, which is the independence of the dispersion from the sign of $a_s$ when $|a_s/a|$ becomes large.  For non-resonant and negative $a_s/a$, picturing the bound states as Fermi pairs with twice the mass and twice the polarizability captures the relative spacings between energy levels quite accurately, but generally \emph{overestimates} the effective mass of the bound states.  This effective mass difference is an indication of the coupling between the center of mass and relative motion which leads to important properties of the FRH.

\emph{Fermi Resonance Hamiltonian.}   A promising route to describing Feshbach interacting ultracold gases is by a lattice projection of a two-channel model in which the closed channel appears explicitly in the Hamiltonian.  However, for a typical broad resonance such models require a large number of both open and closed channel bands to solve accurately, and so cannot be treated efficiently.  Because the modern context of this problem involves extremely low temperatures and densities, we can look for an effective model valid in these limits which still reproduces the correct physics.  This is done by replacing the model containing couplings between all open channel bands with all closed channel bands with a model describing an effective resonance between the lowest open channel band with a suitable set of effective closed channel bands whose properties are set by the two-body solution for low densities.  This process is displayed schematically in Fig.~\ref{fig:EMT}.  The purpose of this section is to derive such an effective Hamiltonian using our two-particle theory.

We begin by separating our two-particle Hamiltonian using projectors $\mathbb{L}$ into the lowest open channel band and $\mathbb{D}=1-\mathbb{L}$ into all excited open channel bands and all closed channel bands.  A similar approach was taken in Ref.~\cite{von_Stecher_Gurarie_11} for the 1D case.  An analysis analogous to that leading to Eqs.~\eqref{eq:NLEE}-\eqref{eq:NLEE2} gives a nonlinear eigenequation for the closed channel components of $\mathbb{D}|\psi\rangle$ as
\begin{align}
\label{eq:NLDEE}&\left[E_{\mathbf{K}}-\nu-{E}_{\mathbf{s}\mathbf{K}}^M\right]R_{\mathbf{s}}^{\mathbf{K}}=\textstyle{\frac{g^2}{a^3}\sum_{\mathbf{t}}}\tilde{\chi}_{\mathbf{st}}^{\mathbf{K}}\left(E_{\mathbf{K}}\right)R_{\mathbf{t}}^{\mathbf{K}}\, ,\\
&\tilde{\chi}_{\mathbf{st}}^{\mathbf{K}}\left(E_{\mathbf{K}}\right)\equiv\textstyle{\sum_{\mathbf{mn};\mathbf{q}}'}\frac{h_{\mathbf{s}\mathbf{K}}^{\mathbf{nm}}\left(\mathbf{q}\right) {h_{\mathbf{t}\mathbf{K}}^{\mathbf{nm}}}^{\star}\left(\mathbf{q}\right)}{E_{\mathbf{K}}-E_{\mathbf{nm}}^{\mathbf{K}}\left(\mathbf{q}\right)+i\eta}-\bar{\chi}_{\mathbf{st}}^{\mathbf{K}}\, ,
\end{align}
where the prime on the sum indicates $\left(\mathbf{m},\mathbf{n}\right)\ne \left(\mathbf{1},\mathbf{1}\right)$.  Here $\tilde{\chi}$ differs from $\chi$ in Eqs.~(\ref{eq:NLEE})-(\ref{eq:NLEE2}) in that the summation excludes the lowest band.  We emphasize that the renormalization $\bar{\chi}$ includes all bands and so the detuning and scattering length used in this projected model are those of the full (non-projected) and properly renormalized two-body problem.  We call the eigenstates of this projected system \emph{dressed molecules}.  Here we label distinct eigenstates of Eq.~\eqref{eq:NLDEE} by the parameter $\alpha$.  These solutions share many features of the full solution presented above.  However, the divergence of the $s$-wave scattering length for the lowest energy completely even parity state occurs near $E_{\mathbf{K}}=0$, indicating that scattering resonances in the lowest open channel band are generated by coupling to this state.

We now assume that, at low temperatures and low densities, two particles which are separated by a distance large compared to the effective range of the potential will remain in the lowest band to minimize their energy.  When two particles come together they interact strongly and populate many of the excited open channel bands as well as the closed channel bands.  Because it is rare for more than two particles to come together, the particular populations of the excited states are fixed by the two-particle solution.  The dressed molecules encapsulate the short distance, high energy physics and couple it to the long wavelength, low energy physics of the lowest band fermions through the Feshbach coupling.  The point of connection between the few- and many-body physics is the two-particle scattering length (equivalently $g$ and $\nu$ for narrow resonances), which appears as a parameter in the equation Eq.~\eqref{eq:NLDEE} defining the dressed molecules.

The FRH is
\begin{align}
 \nonumber&\hat{H}_{\mathrm{eff}}=-t_f\textstyle{\sum_{\sigma\in\left\{\uparrow,\downarrow\right\}}\sum_{\langle i,j\rangle}}\hat{a}_{i\sigma}^{\dagger}\hat{a}_{j\sigma}+E_0\textstyle{\sum_{\sigma\in\left\{\uparrow,\downarrow\right\}}\sum_i} \hat{n}_{i\sigma}^{\left(f\right)}\\
 &\nonumber-\textstyle{\sum_{\alpha\in\mathcal{M}}\sum_{i,j}}t^{\alpha}_{i,j}\hat{d}_{i,\alpha}^{\dagger}\hat{d}_{j,\alpha}+\textstyle{\sum_{\alpha\in\mathcal{M}}}\bar{\nu}_{\alpha}\textstyle{\sum_i}\hat{n}_{i\alpha}^{\left(b\right)}\\
\label{eq:effHami}&+\textstyle{\sum_{\alpha\in\mathcal{M}}\sum_{ijk}}g_{i-k,k-j}^{\alpha}\left[\hat{d}_{i,\alpha}^{\dagger}\hat{a}_{j,\uparrow}\hat{a}_{k,\downarrow}+\mathrm{h.c.}\right]\, ,
\end{align}
where $\hat{a}_{i\sigma}^{\dagger}$ creates a particle with spin $\sigma$ in the lowest open channel band Wannier state centered at lattice site $i$, $w_i\left(\mathbf{x}\right)$; $\hat{d}_{i,\alpha}^{\dagger}$ creates a particle in the $\alpha^{th}$ dressed molecule Wannier state centered at site $i$, $\mathcal{W}_{i,\alpha}\left(\mathbf{x},\mathbf{y}\right)$; $\hat{n}_{i\sigma}^{\left(f\right)}$ is the number operator for fermions in the lowest Bloch band; and $\hat{n}_{i\alpha}^{\left(b\right)}$ is the number operator for the $\alpha^{th}$ dressed molecule state.  The set of dressed molecules $\mathcal{M}$ which are included dynamically can be determined on energetic and symmetry grounds from the two-particle solution.  At low energies, only the completely even parity dressed molecule in the lowest sheet~\cite{SupplMat} is relevant to the set $\mathcal{M}$, as all others either have vanishing on-site couplings from parity considerations or are very far off-resonance.  In order, the terms in Eq.~(\ref{eq:effHami}) represent tunneling of atoms in the lowest Bloch band between neighboring lattice sites $i$ and $j$; the energy $E_0=\sum_{\mathbf{q}}E_{\mathbf{1},\mathbf{q}}/N^3$ of a fermion in the lowest band with respect to the zero of energy; tunneling of the dressed molecular center of mass between two lattice sites $i$ and $j$, not necessarily nearest neighbors; detunings of the dressed molecules from the lowest band two-particle scattering continuum; and resonant coupling between the lowest band fermions at sites $j$ and $k$ in different internal states and a dressed molecule at site $i$.  The FRH is a two-channel resonance model, between unpaired fermions in the lowest band, and dressed molecules nearby in energy.

\begin{figure}[t]
\centerline{\includegraphics[width=\columnwidth]{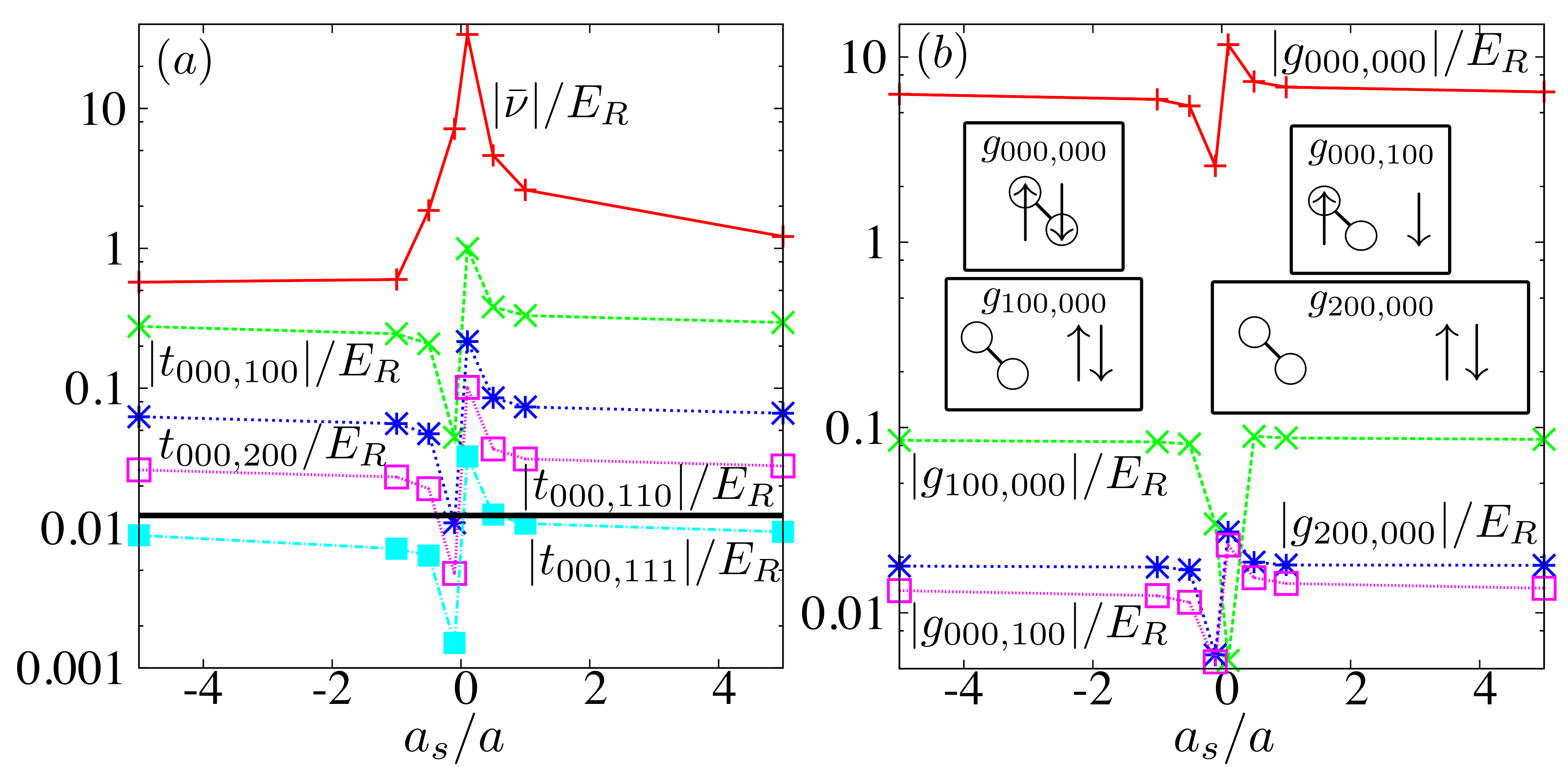}}
\caption{\label{fig:effHstuff} (Color online)  \emph{Hubbard parameters for the FRH}.  (a) The detunings and tunnelings of the completely even parity dressed molecule in the lowest sheet as a function of $a_s/a$.  The detuning is negative for $a_s<0$ and positive otherwise.  The solid black horizontal line is the tunneling of a single open channel fermion in the lowest band.  The nearest neighbor dressed molecular tunneling is nearly two orders of magnitude larger than the open channel tunneling near resonance.  (b) The effective atom-dressed molecule couplings  of the completely even parity dressed molecule in the lowest sheet as a function of $a_s/a$.  Schematics of the spatial dependence of the various coupling processes are shown in the boxes.}
\vspace{-0.2in}
\end{figure}

We now describe how to calculate the Hubbard parameters appearing in Eq.~\eqref{eq:effHami}.  The first term is well-known from single-band Hubbard models~\cite{jaksch_bruder_98} and we do not discuss it here.  Due to the fact that the solutions of the projected nonlinear eigenequation Eq.~\eqref{eq:NLDEE} are also eigenstates of the total quasimomentum, the second and third terms may be written as $\bar{\nu}_{\alpha}=\textstyle{\int d\mathbf{K}}E_{\mathbf{K}}^{\alpha}/v_0$ and $t_{i,j}^{\alpha}=-\textstyle{\int d\mathbf{K}}e^{i\mathbf{K}\cdot\left(\mathbf{R}_i-\mathbf{R}_j\right)}E_{\mathbf{K}}^{\alpha}/v_0$.  Because the band structure is not separable, $E_{\mathbf{K}}\ne\sum_{i=\left\{x,y,z\right\}}E_{K_i}$, dressed molecules can tunnel along directions which are not the principal axes of the lattice.  This is in stark contrast to single-particle tunneling in Bravais lattices which always occurs along the principal axes.  Thus \emph{diagonal hopping} is a key feature neglected in previous approaches.  In Fig.~\eqref{fig:effHstuff}(a) we show that diagonal hopping is often of the same order of magnitude as the tunneling of open channel fermions in the lowest band.  The signs and magnitudes of the tunnelings and particularly the dressed-molecule atom couplings are crucially affected by the parities of the dressed molecular Wannier functions.  We stress that only a full lattice solution can reproduce these important properties of the Hubbard parameters; the frequently used harmonic oscillator approximation will fail even qualitatively to do so.

%

The remaining Hubbard parameter is the dressed molecule-atom coupling, which becomes in the limit of a broad resonance $g/E_Ra^{3/2}\to \infty$
\begin{align}
g_{i-k,k-j}^{\alpha}&=\textstyle{\sum_{\mathbf{s}}\int \frac{d\mathbf{K}}{v_0}}R^{\mathbf{K}}_{\alpha\mathbf{s}}g_{\alpha\mathbf{K}}\textstyle{\int \frac{d\mathbf{q}}{v_0}}e^{i(\mathbf{K}\cdot\mathbf{R}_{ik}+\mathbf{q}\cdot\mathbf{R}_{kj})}h_{\mathbf{sK}}^{\mathbf{11}}(\mathbf{q})
\end{align}
where the renormalized coupling is $g_{\alpha\mathbf{K}}=E_R/\sqrt{-\mathbf{R}^{\mathbf{K}}_{\alpha}\cdot\left(\partial\tilde{\chi}_{\mathbf{K}}/\partial E_{\mathbf{K}}^{\alpha}\right)\cdot\mathbf{R}^{\mathbf{K}}_{\alpha}}$ and $\mathbf{R}_{ij}=\mathbf{R}_i-\mathbf{R}_j$.  We emphasize that $g_{j,k}^{\alpha}$ has only implicit dependence on the divergent parameter $g/E_Ra^{3/2}$ through $g_{\alpha\mathbf{K}}$ and so remains finite, see Fig.~\ref{fig:effHstuff}(b).  As $g_{\alpha\mathbf{K}}\ll g/a^{3/2}$, the transformation to the FRH has the effect of narrowing the resonance.  In Fig.~\ref{fig:effHstuff}(b) we also see that the on-site coupling $g_{000,000}$ is the dominant energy scale of the problem for large $a_s/a$, and that off-site couplings can also be large compared to other Hubbard parameters such as the open channel tunneling.  Atoms which do not lie along a principal axis can pair to form a molecule, but this effect is much weaker than diagonal tunneling for the completely even parity dressed molecule.

In the derivation of the FRH we use only the bound states of the projected problem and neglect scattering states in higher bands.  This captures the scattering states in the lowest band and nearby bound states, but will fail to capture the physics at higher two-particle relative energy where scattering states in higher bands can play a role.  In order to accommodate these scattering states, one can project out higher bands from $\chi$ as was done for the lowest band, and then include these bands dynamically in the many-body Hamiltonian with renormalized couplings.  In this way, the energetic domain of application of the FRH can be extended arbitrarily at the expense of more dynamical fields.  Within the confines of the two-channel model and the constraint of low energies, the FRH is an accurate representation of the many-body Hamiltonian.  However, intrinsic three-body processes which are not captured by the two-channel model play a role at higher density and lead to corrections to the FRH.  A discussion of these three-body processes is outside the scope of this paper.

In summary, we have studied the bound state properties of two Feshbach interacting fermions in an optical lattice at a range of scattering lengths and quasimomenta.  The bound states of a projected Hamiltonian were used to identify a numerically tractable, efficient Hamiltonian for a low density many-body collection of lattice fermions at arbitrary scattering length and low energies, the Fermi Resonance Hamiltonian.  Our results provide the appropriate starting point for future investigations of strongly interacting lattice fermions.

We acknowledge useful discussions with J.~L.~Bohn, H.~P.~B\"{u}chler, C.~W.~Clark, D.~E.~Schirmer, D.~M.~Wood, and Zhigang Wu.  We also thank H.~P.~B\"{u}chler for providing computer code for comparison. This work was supported by the Alexander von Humboldt Foundation, AFOSR, NSF, and GECO.

\bibliographystyle{prsty}

\begin{center}
 \textbf{\large Supplementary material to ``Microscopic model for Feshbach interacting fermions in an optical lattice with arbitrary scattering length and resonance width''} \\[1em]
 M. L. Wall and L. D. Carr \\[2em]
\end{center}

\section{Derivation of the nonlinear eigenequation.}
Here we review the derivation of Eq.~2 of the main text for the bound states of two fermions in an optical lattice interacting via a zero-range Feshbach resonance.  All quantities have their same meaning as in the main text.  The starting point of our analysis is a two channel model with the open channel spanned by states of two Fermions in different internal states with equal mass $m$ and the closed channel spanned by molecular states with twice the Fermionic mass and twice the Fermionic polarizability.  We describe their interaction via an inter-channel coupling $g$ which couples the pair of open channel Fermions to a closed channel molecule at the center of mass and a detuning $\nu$ between the two channels.  This gives rise to the scattering amplitude
\begin{align}
f(\mathbf{k})&=-\frac{1}{1/a_s+ik+r_bk^2}\, ,
\end{align}
with $s$-wave scattering length $a_s=-2\mu g^2/4\pi\hbar^2\nu$ and effective range $r_B=\pi\hbar^4/\mu^2g^2$.  Here $\mu$ is the reduced mass and $\mathbf{k}$ the incident wavevector.

Denoting the wave function of the two Fermions in the open channel as $\psi(\mathbf{x},\mathbf{y})$ and the wave function of the closed channel molecules as $\phi(\mathbf{z})$, the two-channel Schr\"odinger equation in position representation is
\begin{align}
\nonumber&[E-\hat{H}_0(\mathbf{x})-\hat{H}_0(\mathbf{y})]\psi(\mathbf{x},\mathbf{y})=g\textstyle{\int }d\mathbf{z}\alpha(\mathbf{r})\phi(\mathbf{z})\delta(\mathbf{z}-\mathbf{R})\, ,\\
\nonumber&[E-\nu_0-\hat{H}_0^M(\mathbf{z})]\phi(\mathbf{z})=g\textstyle{\int} d\mathbf{x}d\mathbf{y}\alpha(\mathbf{r})\psi(\mathbf{x},\mathbf{y})\delta(\mathbf{z}-\mathbf{R})\, .
\end{align}
In this expression $\hat{H}_0(\mathbf{x})=-\frac{\hbar^2}{2m}\nabla_{\mathbf{x}}^2+V(\mathbf{x})$ is the single particle Hamiltonian for the open channel and $\hat{H}_0^M(\mathbf{z})=-\frac{\hbar^2}{4m}\nabla_{\mathbf{z}}^2+2V(\mathbf{z})$ is the single particle Hamiltonian for the closed channel.  The subscript $0$ in $\nu_0$ denotes that this is a bare detuning entering the microscopic theory which is related to the physically observable detuning $\nu$ in the limit as the regularization cutoff $\Lambda\to \infty$.  Additionally, we note that the Feshbach regularization $\alpha(\mathbf{r})\to\delta(\mathbf{r})$ in the limit as $\Lambda\to \infty$, where $\delta(\mathbf{r})$ is the Dirac delta function.

The open channel solution with total quasimomentum $\mathbf{K}$ may be parameterized as
\begin{align}
\label{eq:openchannel}\psi_{\mathbf{K}}(\mathbf{x},\mathbf{y})&=\frac{1}{\sqrt{N^3}}\sum_{\mathbf{n}\mathbf{m}}\sum_{\mathbf{q}}\varphi_{\mathbf{nm}}^{\mathbf{K}\mathbf{q}}\psi_{\mathbf{n},\mathbf{q}}(\mathbf{x})\psi_{\mathbf{m},\mathbf{K}-\mathbf{q}}(\mathbf{y})\, ,
\end{align}
where $N^3$ is the total number of unit cells in a 3D lattice with periodic boundary conditions and $\psi_{\mathbf{nq}}(\mathbf{x})$ is a Bloch eigenfunction of the single-particle Hamiltonian.  As in the main text, quantities denoted in bold represent three-component vectors, e.g.~$\mathbf{n}=(n_x,n_y,n_z)$.  Similarly, we parameterize the closed channel wave function as a sum over Bloch states computed for twice the mass and twice the polarizability $\phi_{\mathbf{sK}}(\mathbf{z})$ as
\begin{align}
\label{eq:molexpansion}\phi_{\mathbf{K}}(\mathbf{z})&=\sum_{\mathbf{s}}R_{\mathbf{s}}^{\mathbf{K}}\phi_{\mathbf{s}\mathbf{K}}(\mathbf{z})\, .
\end{align}
Inserting these expansions into the two-channel Schr\"{o}dinger equation yields
\begin{align}
\nonumber&\left[E_{\mathbf{K}}-E_{\mathbf{nm}}^{\mathbf{K}}(\mathbf{q})\right]\varphi_{\mathbf{nm}}^{\mathbf{K}\mathbf{q}}=\frac{g}{\sqrt{a^3}}\sum_{\mathbf{s}}h_{\mathbf{s}\mathbf{K}}^{\mathbf{nm}}(\mathbf{q})R_{\mathbf{s}}^{\mathbf{K}}\, ,\\
\nonumber&\left[E_{\mathbf{K}}-\nu_0-E^M_{\mathbf{s}\mathbf{K}}\right]R_{\mathbf{s}}^{\mathbf{K}}=\frac{g}{\sqrt{a^3}}\sum_{\mathbf{n}\mathbf{m}}\int\frac{d\mathbf{q}}{v_0}{h_{\mathbf{s}\mathbf{K}}^{\mathbf{nm}}}^{\star}(\mathbf{q})\varphi_{\mathbf{nm}}^{\mathbf{K}\mathbf{q}}\, .
\end{align}
Formally solving the first of the two equations  with a Green's function and inserting into the second equation gives 
\begin{align}
\nonumber&\left[E_{\mathbf{K}}-\nu_0-{E}^M_{\mathbf{s}\mathbf{K}}\right]R_{\mathbf{s}}^{\mathbf{K}}=\frac{g^2}{a^3}\textstyle{\int\frac{d\mathbf{q}}{v_0}\sum_{\mathbf{mnt}}}\frac{h_{\mathbf{s}\mathbf{K}}^{\mathbf{nm}}\left(\mathbf{q}\right) {h_{\mathbf{t}\mathbf{K}}^{\mathbf{nm}}}^{\star}\left(\mathbf{q}\right)}{E_{\mathbf{K}}-E_{\mathbf{nm}}^{\mathbf{K}}\left(\mathbf{q}\right)+i\eta}R_{\mathbf{t}}^{\mathbf{K}}\, .
\end{align}
This expression diverges in the limit $\Lambda\to \infty$, as is well known for two-channel theories involving a pointlike boson~\cite{Kokkelmans_Milstein_02}.  We remove this divergence through renormalization, replacing the bare detuning $\nu_0$ with the physical detuning $\nu$ by subtracting the infinite constant $\bar{\chi}^{\mathbf{K}}$, yielding Eq.~2 of the main text.  The divergent parts of Eq.~3 in the main text cancel and we may safely take the limit $\Lambda\to \infty$.

Following Ref.~\cite{Buechler_12}, we use the regularization
\begin{align}
\alpha(\mathbf{r})&=\int_{v(\Lambda)}\frac{d\mathbf{k}}{(2\pi)^3}e^{i\mathbf{k}\cdot\mathbf{r}}\, ,
\end{align}
where the cubical volume $v(\Lambda)=v_0\Lambda^3$ is centered around $\mathbf{k}=0$ with $v_0$ the volume of the BZ.  We also define a shell summation over bands with shell parameter $S$, $\sum_{\mathbf{nm};S}$, as the summation over all band indices $\mathbf{n}$ and $\mathbf{m}$ less than or equal to $S$ with at least one of the band indices being $S$.  The correct limiting procedure to obtain $\chi_{\mathbf{st}}^{\mathbf{K}}(\mathbf{K})$ in the limit of an infinite summation over bands and vanishing short-distance cutoff is
\begin{align}
\lim_{\Lambda\to \infty}\left[\lim_{S\to\infty}\chi_{\mathbf{st}}^{\mathbf{K}}(E_{\mathbf{K}})\right]\, .
\end{align}
The $\Lambda$ limit is taken using the asymptotic scaling relation
\begin{align}
\label{eq:chiscaling}\left[\chi_{\mathbf{st}}^{\mathbf{K}}(E_{\mathbf{K}})\right](\Lambda)&=c_{\mathbf{st}}/\Lambda+\chi_{\mathbf{st}}^{\mathbf{K}}(E_{\mathbf{K}})\, .
\end{align}

One may be concerned that the scaling relation Eq.~\eqref{eq:chiscaling} only holds for $\chi^{\mathbf{K}}$ of the full model and not for $\tilde{\chi}^{\mathbf{K}}$ in the projected model.  To show that this is not the case, we note that in the limit of an infinite number of unit cells $N\to\infty$ the overlaps $h_{\mathbf{sK}}^{\mathbf{nm}}(q)$ may be written as products of 1D overlaps $h_{sK}^{nm}(q)$ of the form
\begin{align}
\label{eq:hinf}h_{sK}^{nm}(q)=&\sum_{r,r'}c_{nq}^{r}c_{mK-q}^{r'}{c}_{M;sK}^{r+r'+f/2\pi}\\
\nonumber &\times\mathrm{rect}(\frac{2q-K-f+2\pi(r-r')}{2\pi \Lambda})\, ,
\end{align}
where $f$ is an integer multiple of $2\pi$ which shifts $K-q$ into the BZ, $\mathrm{rect}(x)$ denotes the rectangle function, and the vectors $\mathbf{c}_{nq}$ denote the Fourier expansion of the open channel Bloch functions as
\begin{align}
\psi_{nq}(x)&=e^{i q x}\sum_r c_{nq}^re^{-2\pi i r x/a}/\sqrt{Na}\, .
\end{align}
Similarly, ${\mathbf{c}}_{M;sK}$ denote the Fourier coefficients of the closed channel Bloch functions.  $\tilde{\chi}^{\mathbf{K}}$ differs from $\chi^{\mathbf{K}}$ in the exclusion of all terms with $n=m=1$.  However, provided that $\Lambda$ is large enough to capture the support of the vectors $\mathbf{c}_{nq}$ with $n=1$, Eq.~\eqref{eq:hinf} demonstrates that these terms are no longer functions of $\Lambda$.  Thus, the scaling relation Eq.~\eqref{eq:chiscaling} also holds for $\tilde{\chi}^{\mathbf{K}}$.  Similar arguments show that the same scaling holds for $\tilde{\chi}^{\mathbf{K}}$ when any finite number of open channel bands have been projected out.

\section{Classification of the two-particle bound states}
For the simple cubic lattice we consider the Hamiltonian is invariant under reflection in any Cartesian direction: $H(\theta_{\mathcal{R}}\mathbf{x})=H(\mathbf{x}')=H(\mathbf{x})$ where $\mathbf{x}'\equiv\theta_{\mathcal{R}}\mathbf{x}$ is related to $\mathbf{x}$ by changing the sign of all coordinates in some set $\mathcal{R}$: $x_j\to -x_j$, $j\in\mathcal{R}$.  Because the generators of reflection and translation do not commute we cannot find simultaneous eigenfunctions except at high-symmetry points of the BZ. However, the fact that the Hamiltoninan commutes with both operators implies that parity transformations yield relationships between degenerate sets of Bloch functions.  In particular, for the given lattice potential, the invariance under the reflection symmetry generated by $\theta_{\mathcal{R}}$ implies that the Bloch functions transform as
\begin{align}
\label{eq:inversion}\psi_{\mathbf{nq}}(\mathbf{x}')&=\prod_{j\in\mathcal{R}}(-1)^{n_j+1} \psi_{\mathbf{n},\mathbf{q}'}(\mathbf{x})\, ,
\end{align}
where we begin indexing the bands from 1.  We can thus characterize the bands according to whether they are even or odd under inversions by the triple $\mathbf{p}=(p_x,p_y,p_z)$, where $p_{\nu}=(-1)^{n_{\nu}+1}$. This inversion relationship implies that the inter-channel overlaps transform as
\begin{align}
\label{eq:htransform}h_{\mathbf{s}\mathbf{K}'}^{\mathbf{nm}}(\mathbf{q}')&=\prod_{j\in\mathcal{R}}(-1)^{n_j+m_j+s_j+1}h_{\mathbf{s}\mathbf{K}}^{\mathbf{nm}}(\mathbf{q})\, ,
\end{align}
and $\chi^{\mathbf{K}}$ transforms as
\begin{align}
\label{eq:chitransform}\chi_{\mathbf{st}}^{\mathbf{K}'}(E_{\mathbf{K}})&=\prod_{j\in\mathcal{R}}(-1)^{s_j+t_j}\chi_{\mathbf{st}}^{\mathbf{K}}(E_{\mathbf{K}})\, .
\end{align}
It can be proven that this transformation leaves the eigenvalues invariant, but the eigenvectors $\mathbf{R}^{\mathbf{K}}_{\alpha}$ transform according to
\begin{align}
R_{\mathbf{s}\alpha}^{\mathbf{K}'}&=\prod_{j\in\mathcal{R}}(-1)^{s_j+1}R_{\mathbf{s}\alpha}^{\mathbf{K}}\, .
\end{align}
For a total quasimomentum $\mathbf{K}$ all of whose components consist of either 0 or $-\pi/a$, this implies that only molecular bands which transform identically under \emph{complete} inversions mix.  Hence, at these exceptional points of the BZ, we can unambiguously determine the parity of the two-particle state $\psi_{\mathbf{K}\alpha}(\mathbf{x},\mathbf{y})$ by the components of its associated eigenvector $\mathbf{R}^{\mathbf{K}}_{\alpha}$.  The parity is then chosen to depend only on the eigenstate index $\alpha$ by requiring that $\mathbf{R}_{\alpha}^{\mathbf{K}}$ is a smooth function of $\mathbf{K}$.  This construction follows that of the 1D case studied by Kohn~\cite{Kohn_59}, which leads to maximally localized Wannier functions.

With this construction, there is still an undefined global phase under inversion that we can fix in the following way.  The complete two-particle bound state solution is
\begin{align}
\label{eq:tpstate}&\Psi_{\mathbf{K}\alpha}(\mathbf{x},\mathbf{y})=\frac{1}{\mathcal{N}_{\mathbf{K}\alpha}}\Big[\sum_{\mathbf{s}}R^{\mathbf{K}}_{\mathbf{s}\alpha}\phi_{\mathbf{sK}}(\mathbf{x})\tilde{r}(\mathbf{x}-\mathbf{y})\\
\nonumber&+\frac{g}{\sqrt{N^3a^3}}\sum_{\mathbf{nms};\mathbf{q}}\frac{R_{\mathbf{s}\alpha}^{\mathbf{K}}h_{\mathbf{sK}}^{\mathbf{nm}}(\mathbf{q})\psi_{\mathbf{nq}}(\mathbf{x})\psi_{\mathbf{mK}-\mathbf{q}}(\mathbf{y})}{E_{\mathbf{K}}^{\alpha}-E_{\mathbf{nm}}^{\mathbf{K}}(\mathbf{q})}\Big]\, ,
\end{align}
where $\mathcal{N}_{\mathbf{K}\alpha}$ is a normalizing factor and $\tilde{r}(\mathbf{x}-\mathbf{y})$ denotes a relative wavefunction for the closed channel which has characteristic width $a/\Lambda$.  As the theory has already been regularized, we may take $\Lambda\to \infty$ with the understanding that this relative wavefunction has a probability density of 1, and forces the closed channel to contribute only at the center of mass.  Because of the partitioning of Hilbert space into open and closed channels, the normalization coefficient is
\begin{align}
\mathcal{N}_{\mathbf{K}\alpha}^2&=1-(\frac{g}{E_Ra^{3/2}})^2\mathbf{R}^{\mathbf{K}}_{\alpha}\cdot\chi'(E_{\mathbf{K}}^{\alpha}/E_R)\cdot\mathbf{R}^{\mathbf{K}}_{\alpha}\, .
\end{align}
Here $\chi'(E)$ is the derivative of $\chi$ with respect to $E$.  Using the transformation properties under ${\theta}_{\mathcal{R}}$, we find
\begin{align}
\Psi_{\mathbf{K}'\alpha}(\mathbf{x},\mathbf{y})
\label{eq:twoparticleinversion}&=P_{\alpha}\Psi_{\mathbf{K}\alpha}(\mathbf{x}',\mathbf{y}')\, .
\end{align}
Accordingly, we set $P_{\alpha}=\prod_{j\in\mathcal{R}}p_{\nu}$.  This implies that the dressed molecular Wannier functions transform as $\mathcal{W}_{i\alpha}(\mathbf{x}',\mathbf{y}')=P_{\alpha}\mathcal{W}_{i\alpha}(\mathbf{x},\mathbf{y})$.

\begin{figure}[t]
\centerline{\includegraphics[width=0.75\columnwidth]{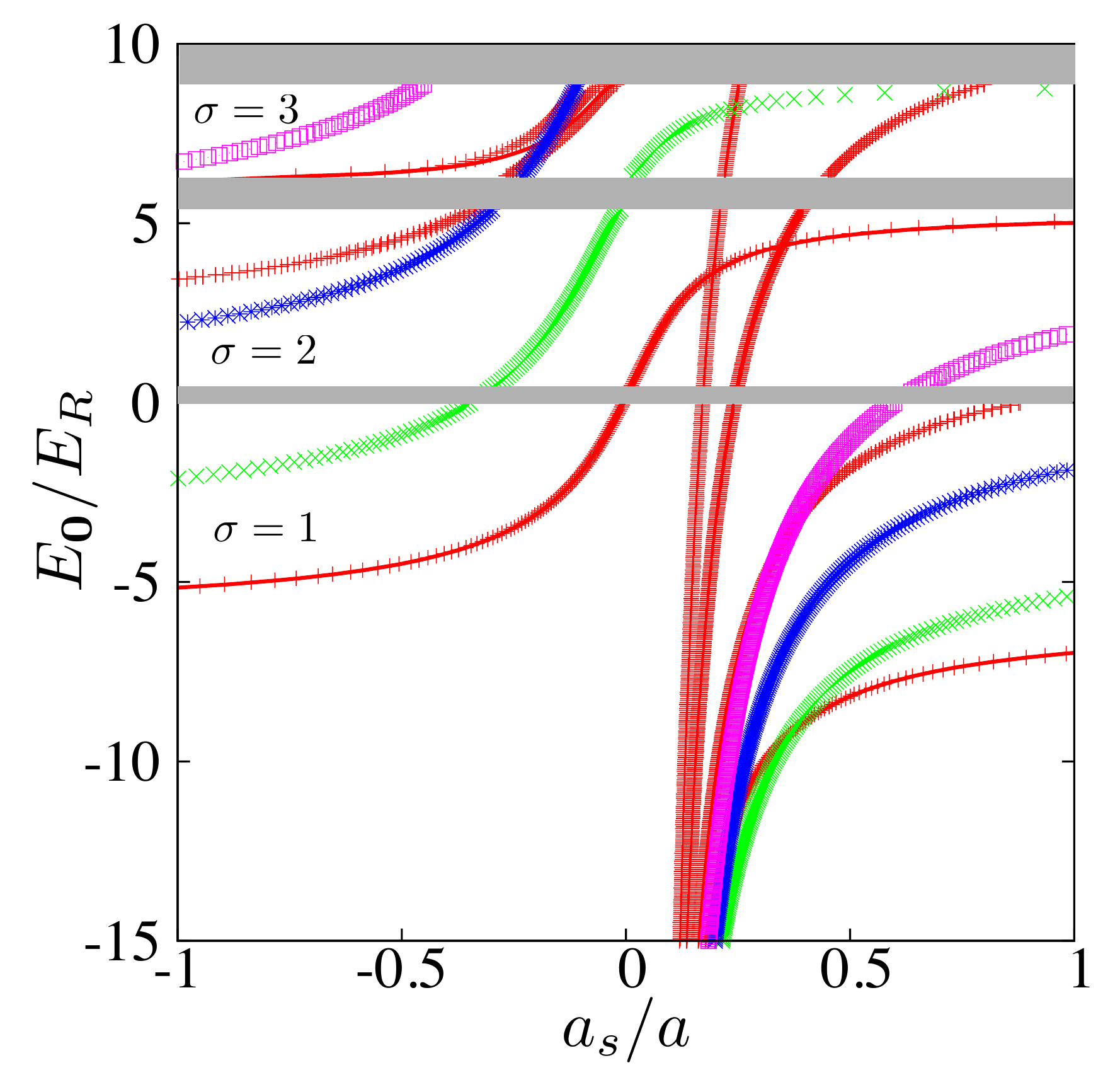}}
\caption{\label{fig:class} (Color online)  \emph{Classification of two-particle bound states}.  The bound state energies at $\mathbf{K}=0$ for a lattice of depth $V/E_R=12$ are classified according to their parity and sheet indices.  Red corresponds to $\mathbf{p}=(1,1,1)$, green to $\mathbf{p}=(1,1,-1)$, blue to $\mathbf{p}=(1,-1,-1)$, and magenta to $\mathbf{p}=(-1,-1,-1)$.  In addition, higher molecular bands for the $\mathbf{p}=(1,1,1)$ level are shown.  These give rise to weak scattering resonances with the lowest open channel band for $a_s/a>0$ and avoided crossings in the higher sheets for $a_s/a<0$.}
\end{figure}

In Fig.~\ref{fig:class} we display the bound state energies at $\mathbf{K}=0$ for a lattice of depth $V/E_R=12$ classified according to their parity.  The red points correspond to $\mathbf{p}=(1,1,1)$, the green points to $\mathbf{p}=(1,1,-1)$ et.~cyc., the blue points to $\mathbf{p}=(1,-1,-1)$ et.~cyc., and the magenta points to $\mathbf{p}=(-1,-1,-1)$.  In contrast to the continuum where there exists at most one bound state for fixed scattering length $a_s$, there is the possibility of several bound states for fixed $a_s$ in the lattice due to the reduced translational symmetry.  Thus, the parity and the quasimomentum are not sufficient to completely describe the states.  For a fixed $s$-wave scattering length $a_s$, we provide two other indices which we call the \emph{sheet index} $\sigma\in\left\{1,2,\dots, \infty\right\}$ and the \emph{molecular band index} $s\in \left\{1,2,\dots,\infty\right\}$.  The sheet index labels the open channel two-particle scattering bands, with the convention that the first sheet lies below the first band, the second sheet lies between the first and second bands, etc. as indicated on the figure.  The open channel scattering bands are denoted by solid stripes.  The molecular band index labels eigenstates which have the same parity and sheet indices but differ in energy.  The number of molecular bands obtained is restricted by the number of closed channel bands used to construct $\chi^{\mathbf{K}}$.  Let us define $m$ to be the maximum value of the closed channel band index along any Cartesian direction.  In Fig.~\ref{fig:class}, the solid red line corresponds to the completely even parity state computed with $m=2$ and the red points correspond to the complete even parity states computed with $m=3$.  $m=2$ captures the physics well near the lowest open channel scattering band.  For $a_s/a>0$, the higher molecular bands cross the lowest open channel scattering continuum at narrow ranges of $a_s/a$, leading to weak scattering resonances.  For $a_s/a<0$, the higher molecular bands are present at extended ranges of $a_s/a$, and avoided crossings between these molecular bands can lead to differences with lower $m$ computations, see e.g.~the third sheet near $a_s/a=-0.2$.

\section{Finite width resonances}
\begin{figure}[t]
\begin{center}
\centerline{\includegraphics[width=0.75\columnwidth]{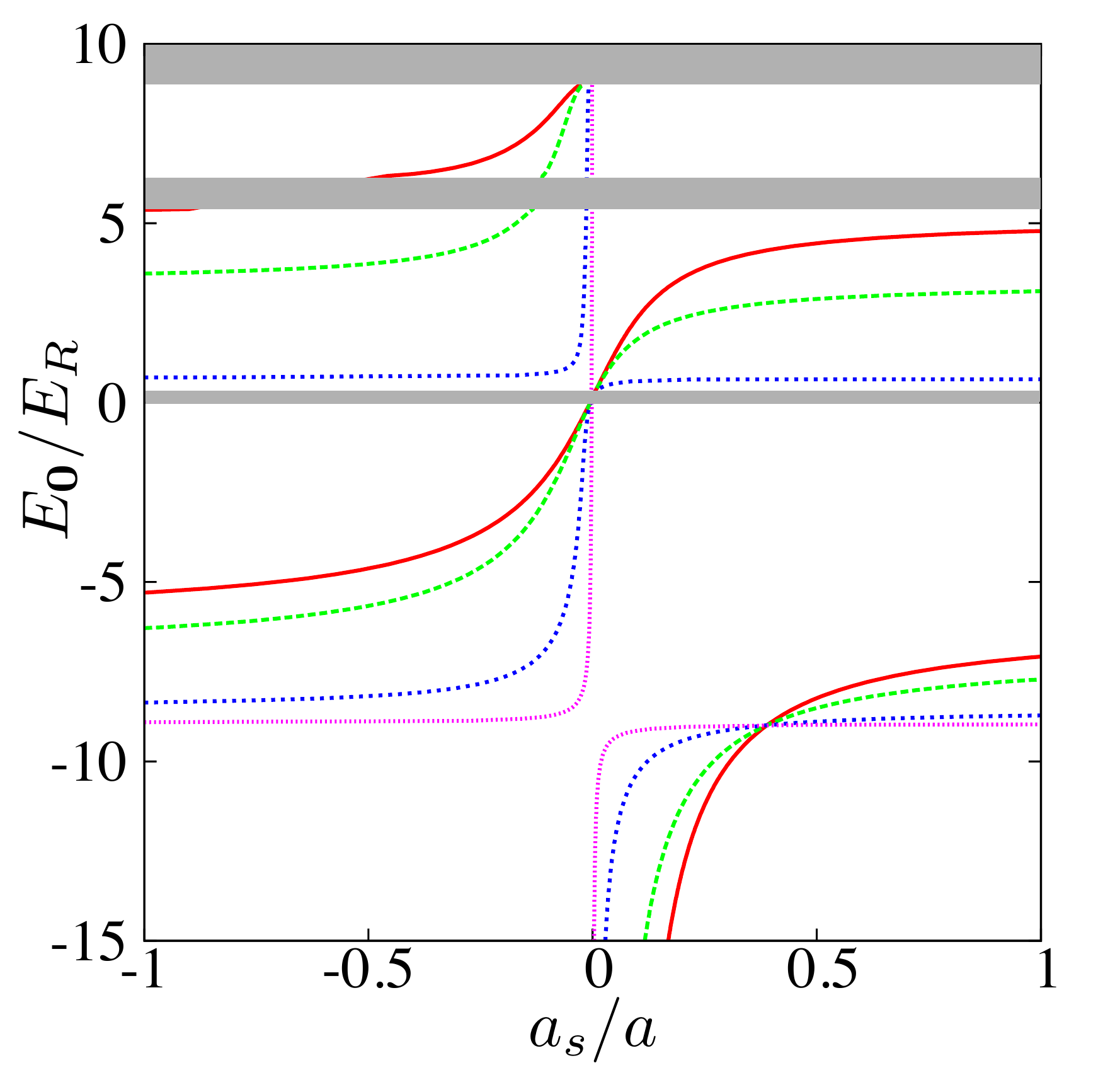}}
\caption{\label{fig:narrow} (Color online)  \emph{Bound state energies for finite width resonances.}  Shown are the lowest energy bound state energies for $\mathbf{p}=(1,1,1)$ at $\mathbf{K}=0$ in a strong optical lattice with $V/E_R=12$.  The red solid line is $r_B/a=0.01$, the green dashed line is $r_B/a=0.1$, the blue dotted line is $r_B/a=1$, and the magenta short-dashed line is $r_B/a=10$.  For narrow resonances (large $r_B$) the divergence of $a_s$ is sharply pronounced around a narrow energy range, and is shifted downwards from the broad resonance value, compare Fig.~\ref{fig:class}.}
\end{center}
\end{figure}

We now turn our attention briefly to the case where $g/E_Ra^{3/2}$ and $\nu/E_R$ are finite.  In this case we rearrange Eq.~2 of the main text to read
\begin{align}
\sum_{\mathbf{t}}\left[(E_{\mathbf{K}}-E_{\mathbf{sK}}^M)\delta_{\mathbf{st}}-\frac{g^2}{a^3}\chi^{\mathbf{K}}_{\mathbf{st}}(E_{\mathbf{K}})\right]R_{\mathbf{t}}^{\mathbf{K}}&=\nu R_{\mathbf{s}}^{\mathbf{K}}
\end{align}
which is an ordinary eigenvalue equation for the detuning $\nu$ when $E_{\mathbf{K}}$ and $g$ are treated as fixed.  We note that $g$ cannot be scaled out of this equation as the molecular band energies $E_{\mathbf{sK}}^M$ depend only on the lattice strengths and masses and not on the resonance width.  The solution of this equation for various $r_B$ and $\mathbf{K}=0$ is shown in Fig.~\ref{fig:narrow}.

We characterize the width of the resonance in terms of the experimentally measurable effective range $r_B$ which defines the width as $g/E_Ra^{3/2}=\sqrt{16a/\pi^3r_B}$.  For narrow resonances with large $r_B$ only the lowest resonance can be seen, and $a_s/a$ is greater than 1, corresponding to strong interactions, only in a very narrow energy range.  As the resonance becomes broader the energy range over which the system is strongly interacting widens, and we begin to see resonant behavior near higher scattering continua.  Additionally, the positions of the narrow resonances are shifted downwards in energy with respect to the broad resonances, eventually becoming the free molecular band energies.  We note that the broadest resonance shown is in fact narrower than typical broad resonances found in the experimentally relevant ultracold atomic systems, but differs from the infinitely broad resonance results by at most a few percent.  This justifies our use of the infinitely broad resonance limit in the main text.

\end{document}